\documentclass[%
reprint,
showpacs,preprintnumbers,
amsmath,amssymb,
aps,
noeprint,
pra,
floatfix
]{revtex4-1}

\usepackage{graphicx}
\usepackage{lineno}
\usepackage{dcolumn}
\usepackage{color}
\usepackage{bm}
\usepackage{multirow}
\usepackage{hyperref}

\begin{document}
\normalsize

\title{Observation of {\boldmath $\eta_c\to\omega\omega$} in {\boldmath $J/\psi\to\gamma\omega\omega$}}

\author{
   \begin{small}
      \begin{center}
         M.~Ablikim$^{1}$, M.~N.~Achasov$^{10,d}$, S. ~Ahmed$^{15}$, M.~Albrecht$^{4}$, M.~Alekseev$^{55A,55C}$, A.~Amoroso$^{55A,55C}$, F.~F.~An$^{1}$, Q.~An$^{52,42}$, Y.~Bai$^{41}$, O.~Bakina$^{27}$
% ...
\vspace{0.2cm}
\\
(BESIII Collaboration)\\
\vspace{0.2cm}
{ \it
$^{1}$ Institute of High Energy Physics, Beijing 100049, People's Republic of China\\
% ...
$^{4}$ Bochum Ruhr-University, D-44780 Bochum, Germany\\
% ...
$^{10}$ G.I. Budker Institute of Nuclear Physics SB RAS (BINP), Novosibirsk 630090, Russia\\
% ...
$^{15}$ Helmholtz Institute Mainz, Johann-Joachim-Becher-Weg 45, D-55099 Mainz, Germany\\
% ...
$^{27}$ Joint Institute for Nuclear Research, 141980 Dubna, Moscow region, Russia\\
% ...
$^{41}$ Southeast University, Nanjing 211100, People's Republic of China\\
$^{42}$ State Key Laboratory of Particle Detection and Electronics, Beijing 100049, Hefei 230026, People's Republic of China\\
% ...
$^{52}$ University of Science and Technology of China, Hefei 230026, People's Republic of China\\
% ...
$^{55A}$ University of Turin, I-10125, Turin, Italy\\
$^{55B}$ University of Eastern Piedmont, I-15121, Alessandria, Italy\\
$^{55C}$ INFN, I-10125, Turin, Italy\\
% ...
$^{d}$ Also at the Novosibirsk State University, Novosibirsk, 630090, Russia\\
% ...
} 
\end{center}
\vspace{0.4cm}
\end{small}
}

\noaffiliation

\begin{abstract}
   Using a sample of $(1310.6\pm7.0)\times10^6$ $J/\psi$ events recorded with the BESIII detector at the symmetric electron positron collider BEPCII, we report the observation of the decay of the $(1^1 S_0)$ charmonium state $\eta_c$ into a pair of $\omega$ mesons in the process $J/\psi\to\gamma\omega\omega$. The branching fraction is measured for the first time to be $\mathcal{B}(\eta_c\to\omega\omega)= (2.88\pm0.10\pm0.46\pm0.68)\times10^{-3}$, where the first uncertainty is statistical, the second systematic and the third is from the uncertainty of $\mathcal{B}(J/\psi\to\gamma\eta_c)$. The mass and width of the $\eta_c$ are determined as $M=(2985.9\pm0.7\pm2.1)\,$MeV/$c^2$ and $\Gamma=(33.8\pm1.6\pm4.1)\,$MeV. 
\end{abstract}


\maketitle

\section{\label{sec:level1}INTRODUCTION}
Although the $\eta_c$ was discovered already in 1980 \cite{Partridge:1980vk}, the properties of the lowest lying \textit{S}-wave spin singlet charmonium state are still under investigation. Especially when considering the available data on the branching fractions (BFs) of different decay modes of the $\eta_c$, it becomes obvious that this resonance is not fully understood yet. Several BFs are only measured very roughly or with large uncertainties, and the observed BFs sum up to only about 57\%. Several peculiarities also arise, when the resonance parameters of this meson are studied in detail: The observed mass and decay width seem to vary by a large fraction from experiment to experiment, and also seem to be dependent on the production, and/or decay process in which they are observed. While the decay of the $\eta_c$ into a pair of $\phi$ mesons has been observed before (see \emph{e.g.} Refs.\,\cite{Ablikim:2005yi}, \cite{Bai:2003tr}), only an upper limit for the decay into two $\omega$ mesons has been set \cite{Baltrusaitis:1985mr}. Apart from these measurements, the Belle experiment was able to determine the product BF $\mathcal{B}(\gamma\gamma\to\eta_c)\times \mathcal{B}(\eta_c\to\omega\omega)$ \cite{belle-omom}. The decay $\eta_c\rightarrow 2(\pi^+\pi^-\pi^0)$, which should also contain a large fraction of the $\omega\omega$ channel, has been determined to be one of the strongest decay modes of the $\eta_c$ \cite{Ablikim:2012ur}.
     Predictions for the BFs of the $\eta_c$ into a pair of vector mesons have been recently published \cite{Sun201149}. 
     However, the predicted BFs for the decay modes $\eta_c\rightarrow\phi\phi$ and $\eta_c\rightarrow\rho\rho$ are much smaller than those observed experimentally. 
     The predictions are based on Next-to-Leading order (NLO) perturbative Quantum Chromodynamics (QCD) calculations and for the first time also include so-called higher-twist contributions. It was found that these contributions do have a major impact on the BFs and lead to much larger values than expected from pure perturbative QCD. However, the effect is not strong enough to explain the experimentally determined BFs for the $\phi\phi$ and $\rho\rho$ channels. 
     The predictions for the BF of the $\eta_c\to\omega\omega$ process range from $9.1\times10^{-5}$ to $1.3\times10^{-4}$, while the most sensitive experimental determination yielded an upper limit of $<3.1\times10^{-3}$ at the $90\%$ confidence level \cite{Baltrusaitis:1985mr}.
      
      In this paper we present the first measurement of the BF for the decay $\eta_c\to\omega\omega$, where the $\eta_c$ is observed in the invariant mass of two $\omega$ mesons produced in the radiative decay $J/\psi\to \gamma\omega\omega$. The data set used for this analysis contains a total of $(1310.6\pm7.0)\times10^6$ $J/\psi$ events \cite{Ablikim:2016fal} produced in direct $e^+e^-$ annihilations and recorded with the Beijing Spectrometer III (BESIII) detector. The mass, the width, and the yield of the $\eta_c$ signal are determined by means of a partial wave analysis (PWA) in the $\eta_c$ signal region to properly account for interference effects with other contributions to the $\omega\omega$ system.

\section{DETECTOR AND MONTE CARLO SIMULATION}
The BESIII detector \cite{besiiiNIM} is located at the electron positron collider BEPCII \cite{bepciiConstr} at the Institute for High Energy Physics (IHEP), Beijing, China. The symmetric double-ring collider BEPCII provides a peak luminosity of $10^{33}$cm$^{-2}$s$^{-1}$ at a center-of-mass energy of $3.77$\,GeV. The detector consists of four main components: A small-cell gas drift chamber with 43 layers directly surrounds the beam pipe. This main drift chamber (MDC) is filled with a 60\% He, 40\% C$_3$H$_8$ gas mixture. It provides an average single-hit position resolution of $135\,\mu$m as well as a charged particle momentum resolution of $0.5\%$ ($0.6\%$) at 1\,GeV$/c$ in a 1\,T (2009) or 0.9\,T (2012) magnetic field, which is generated by a superconducting solenoid magnet. The $dE/dx$ resolution of the MDC is 6\% for electrons from Bhabha scattering. Surrounding the drift chamber, a plastic scintillator based time-of-flight system (TOF) for particle identification followed by a CsI(Tl)-based electromagnetic calorimeter (EMC) is mounted. The EMC consists of 6240 crystals arranged in a cylindrical, barrel-shaped part and two end caps. The calorimeter provides an energy resolution of 2.5\% (5\%) for 1\,GeV photons as well as a position resolution of 6\,mm (9\,mm) in the barrel (end caps). The time-of-flight system consists of 176 scintillator bars with a length of 2.4\,m, arranged in a two-layer, barrel-shaped geometry and 96 fan-shaped scintillators in the end caps. All plastic scintillators of the time-of-flight system have a thickness of 5\,cm. The system provides a $K/\pi$ separation of $2\sigma$ for momenta up to $\sim1\,$GeV$/c$ with a time resolution of 80\,ps (110\,ps) in the barrel (end caps). The iron return yoke of the solenoid magnet is instrumented with 9 (8) layers of resistive plate chambers in the barrel (end cap) regions, yielding in total about $1272$\,m$^2$ of active area. The signals from these chambers can be used for muon identification with a position resolution of 2\,cm.

Phase-space distributed Monte Carlo (MC) data sets of the signal channel are generated for optimizations of the event selection over the complete phase-space (26M events) as well as the minimization in the PWA containing only events in the $\eta_c$ mass range (2M events). The simulations are carried out using a {\sc Geant4}-based simulation software, which includes a precise description of the BESIII geometry and material, the detector response and digitization models, as well as the detector running conditions and performance. The production of the $J/\psi$ resonance is simulated by the MC generator {\sc KKMC} \cite{KKMC}. The subsequent decay of the $J/\psi$ into a radiative photon and a pair of $\omega$ mesons, as well as the three-body decays of the $\omega$ mesons into $\pi^+\pi^-\pi^0$ are generated using {\sc BesEvtGen} \cite{besevtgen}, which is based on the {\sc EvtGen} package \cite{evtgen}.

\section{EVENT SELECTION}
We perform an exclusive reconstruction of the decay $J/\psi\to\gamma\omega\omega$, where both $\omega$ mesons are reconstructed in their decay into $\pi^+\pi^-\pi^0$. Both $\pi^0$ mesons decay further into a pair of photons, thus yielding the final state $\pi^+\pi^-\pi^+\pi^-5\gamma$.
Candidate events are required to contain two pairs of oppositely charged tracks and at least five photon candidates. 
        
Tracks of charged particles are reconstructed using the hit information from the MDC. A track is accepted as a charged particle candidate if the distance between the point of closest approach and the interaction point is smaller than 1\,cm in the plane perpendicular to the beam and smaller than 10\,cm in the beam direction. Furthermore, each track is required to be within the angular acceptance of the MDC, fulfilling the requirement on the polar angle $|\cos \theta|<0.93$. 
    
      Pion candidates are selected from all good charged tracks, by exploiting the capabilities of particle identification of the different subdetector systems. Using the information on the energy loss $dE/dx$ measured with the MDC, as well as the information from the time-of-flight system, a likelihood is calculated under the hypotheses that the particle candidate under investigation is a pion ($\mathcal{L}(\pi)$), kaon ($\mathcal{L}(K)$) or proton ($\mathcal{L}(p)$). Only candidates fulfilling the criteria $\mathcal{L}(\pi)>\mathcal{L}(K)$ and $\mathcal{L}(\pi)>\mathcal{L}(p)$ are accepted and retained for further analysis.

      Photon candidates are showers detected with the EMC exceeding an energy of 25\,MeV in the barrel (${|\cos\theta|<0.8}$) and 50\,MeV in the end cap regions (${0.86 < |\cos\theta| < 0.92}$), respectively. To reject photons originating from split-off effects, each photon candidate must lie outside a cone with an opening angle of $20^\circ$ around the impact point in the calorimeter of any charged track. Furthermore, photon candidates are only accepted if their hit time is within 700\,ns of the event start time to suppress electronic noise and showers that are unrelated to the event.

\begin{figure}[t]
        \includegraphics[width=.4\textwidth]{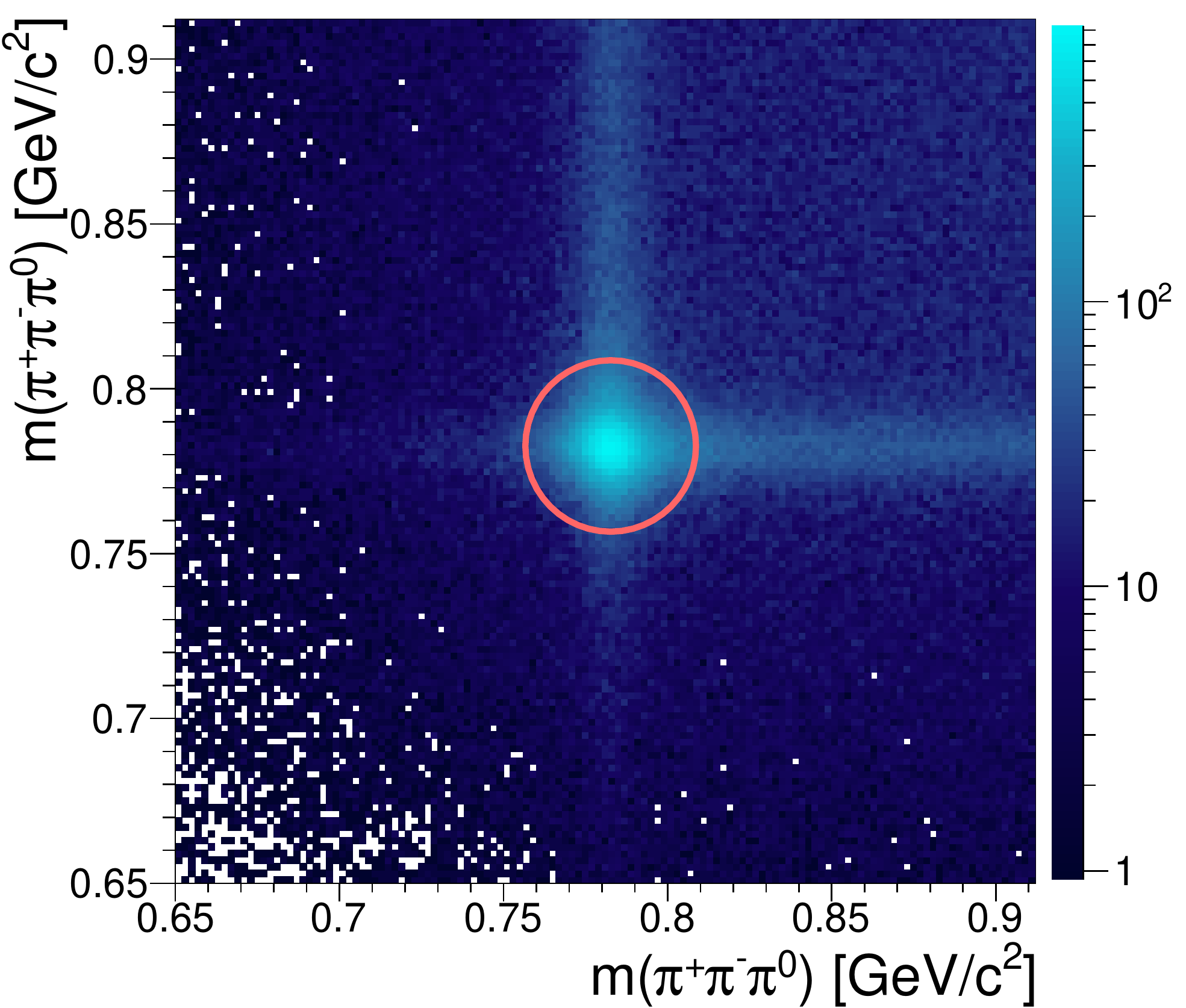}
        \caption{Distribution of the invariant masses of both three-pion systems appearing in the decay $J/\psi\to\gamma (\pi^+\pi^-\pi^0)_1 (\pi^+\pi^-\pi^0)_2$ for the chosen best combination of each event. The bands correspond to the mass of the $\omega$ meson; a clear enhancement at the intersection of the two bands is visible. The red circle indicates the signal region which is selected after application of the background subtraction method described in section \ref{williams}.}
        \label{fig01}
     \end{figure}

     To improve the momentum resolution of the $\omega$ candidates, suppress background and determine the correct combination of photons to form $\pi^0$ candidates, all events are kinematically fitted under the $J/\psi\to\gamma\pi^+\pi^-\pi^0\pi^+\pi^-\pi^0$ hypothesis for all possible combinations of photons.
     The fit is performed using six kinematic constraints, which are the energy and the three linear momentum components of the initial $e^+e^-$ system, as well as the masses of the two $\pi^0$ candidates. The combination which yields the smallest $\chi^2_{6C}$ value for the kinematic fit is chosen and the event is kept for further analysis, if $\chi^2_{6C}<25$. This effectively reduces photon mis-combination. Finally the correct combination of two sets of three pions to form the two $\omega$ candidates must be found. The three pions are assigned to the $\omega$ candidate, for which they exhibit the closest Euclidean distance $r$ from the nominal mass of the $\omega$ meson, given by
     
     \begin{align}
        r=\sqrt{[m(3\pi)_1-m(\omega)]^2 + [m(3\pi)_2-m(\omega)]^2}.
     \end{align}
     Here, $m(\omega)$ indicates the nominal mass of the $\omega$ meson as listed in Ref.\,\cite{pdg2016}.  
     Figure \ref{fig01} shows the $3\pi$ versus $3\pi$ invariant mass for all events retained after the selection procedure described above. 
     
     Two bands originating from the process $J/\psi\to\gamma\omega3\pi$, located at the nominal $\omega$ mass, are clearly visible in Fig. \ref{fig01}. Additionally, a flat, homogeneous background corresponding to $J/\psi\to\gamma6\pi$ events is visible. Events from both of these processes are also present under the clearly visible enhancement at the intersection of the two $\omega$ bands. 
     To remove this type of background, an event-based background subtraction method is used, which is described in the following section. After application of the background subtraction, a strict selection requirement around the intersection of the two bands is introduced.
     
\section{Background subtraction}
\label{williams}
A sophisticated event-based method for background subtraction proposed in Ref.\,\cite{Williams:2008sh} is applied to events for which both three-pion invariant masses are located within a range of $\pm80\,$MeV around the nominal $\omega$ mass. 
Simpler methods, such as a two-dimensional side band subtraction, mostly require the analysis of a binned data set, while the goal here is to perform a PWA and thus an event-based method is preferred. 

The method is based on analyzing the signal-to-background ratio $Q$ in a very small cell of the available phase-space around each event. 
Therefore a distinct kinematic variable is needed, for which parameterizations of both the signal and background shape are known for the events in these small cells.
The first step is to assign a number of $N$ nearest neighbors for each event, denoted as seed event. In order to measure distances between events, a metric has to be defined using the kinematic observables that span the phase space for the reaction. For this analysis, in total nine coordinates are used for the metric: the polar angle of the radiative photon in the $J/\psi$ rest frame, where the $z$-axis is defined by the direction of the incoming positron beam, the angle between the two $\omega$ candidates' decay planes in the $J/\psi$ rest frame, the invariant mass of the $2(\pi^+\pi^-\pi^0)$ system, the azimuthal and polar decay angles of the two $\omega$ candidates in the helicity frame of the corresponding $\omega$ candidate, as well as the two normalized slope parameters $\tilde{\lambda}$ of the $\omega$ candidates' decays. The parameter $\tilde{\lambda}$ characterized by the cross product of the two pion momenta in the corresponding $\omega$ candidates' helicity frame is given as
     \begin{align}
        & \tilde{\lambda} = \lambda' / \lambda'_{\textnormal{max}} \textnormal{ with } \lambda' = |\vec{p}_{\pi^+} \times \vec{p}_{\pi^-}|^2 \label{eqn2}\\
                &\textnormal{and } \lambda'_{\textnormal{max}} = T^2 \left(\frac{T^2}{108c^4} + \frac{m_\pi T}{9c^2} + \frac{m_\pi^2}{3} \right), \nonumber\\ 
                &T = T_{\pi^+} + T_{\pi^-} + T_{\pi^0}, \nonumber
   \end{align}
   \noindent where $T_{\pi}$ denotes the kinetic energy of the corresponding pion \cite{Weidenauer:1993mv} and $c$ is the speed of light. The parameter $\lambda'$ takes its maximum value $\lambda'_\text{max}$ for totally symmetric decays with an angle of $120^\circ$ between any pion pair (see Ref.\cite{Weidenauer:1993mv}). The distance between two events is given by the Euclidean distance considering all coordinates listed above.
    
    \begin{figure}[t]
      \includegraphics[width=.23\textwidth]{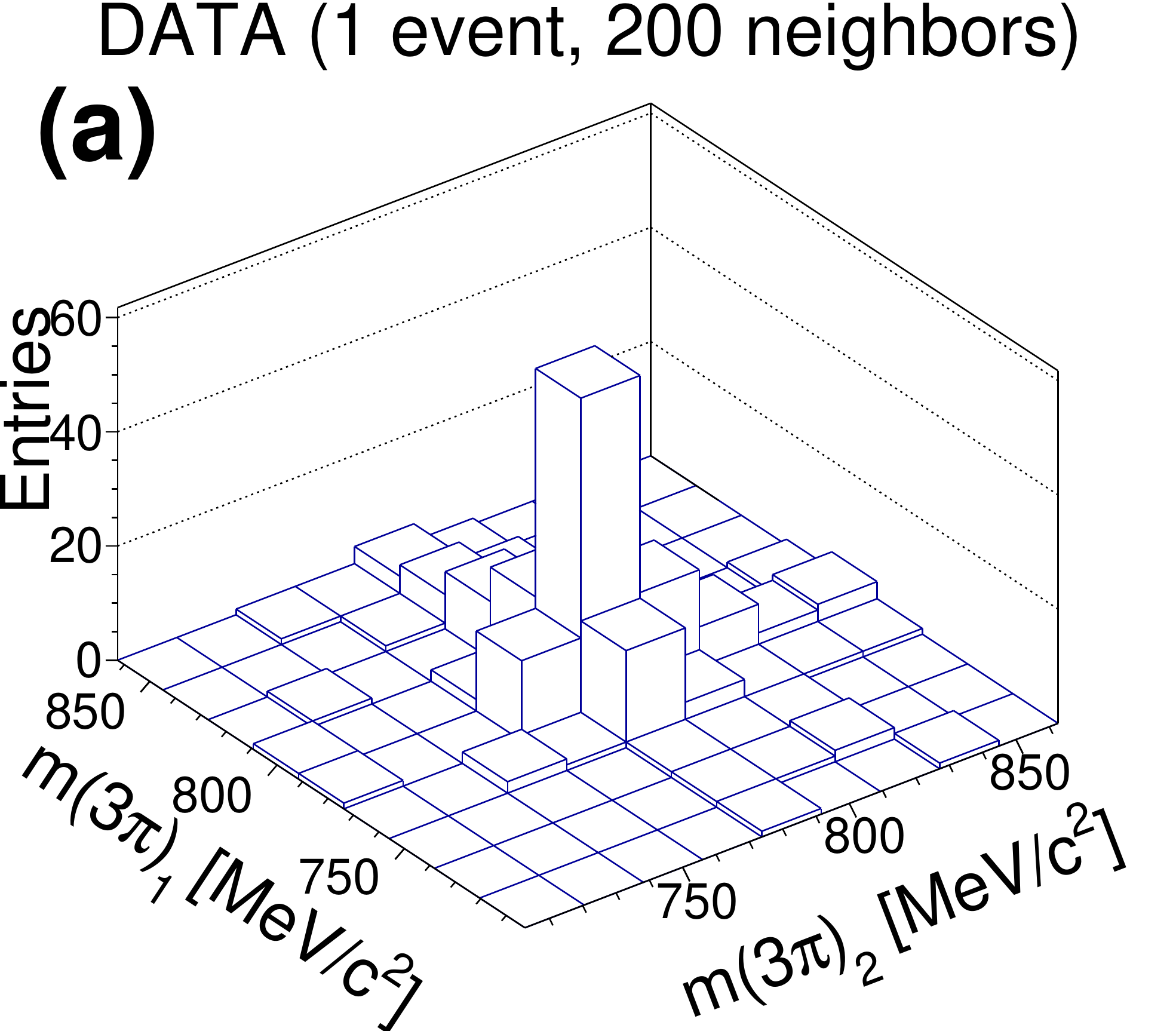} 
      \includegraphics[width=.23\textwidth]{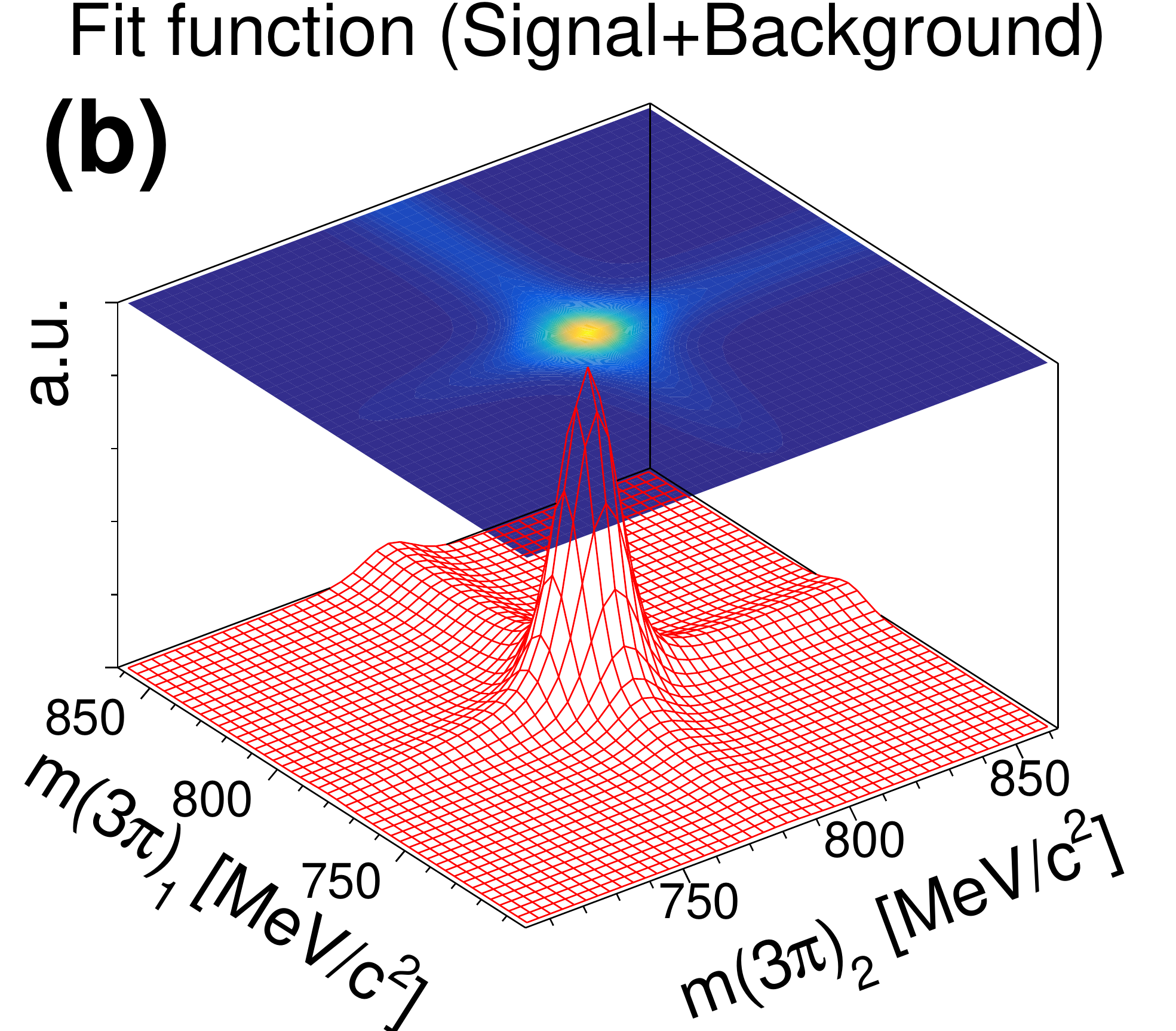} \\
      \includegraphics[width=.22\textwidth]{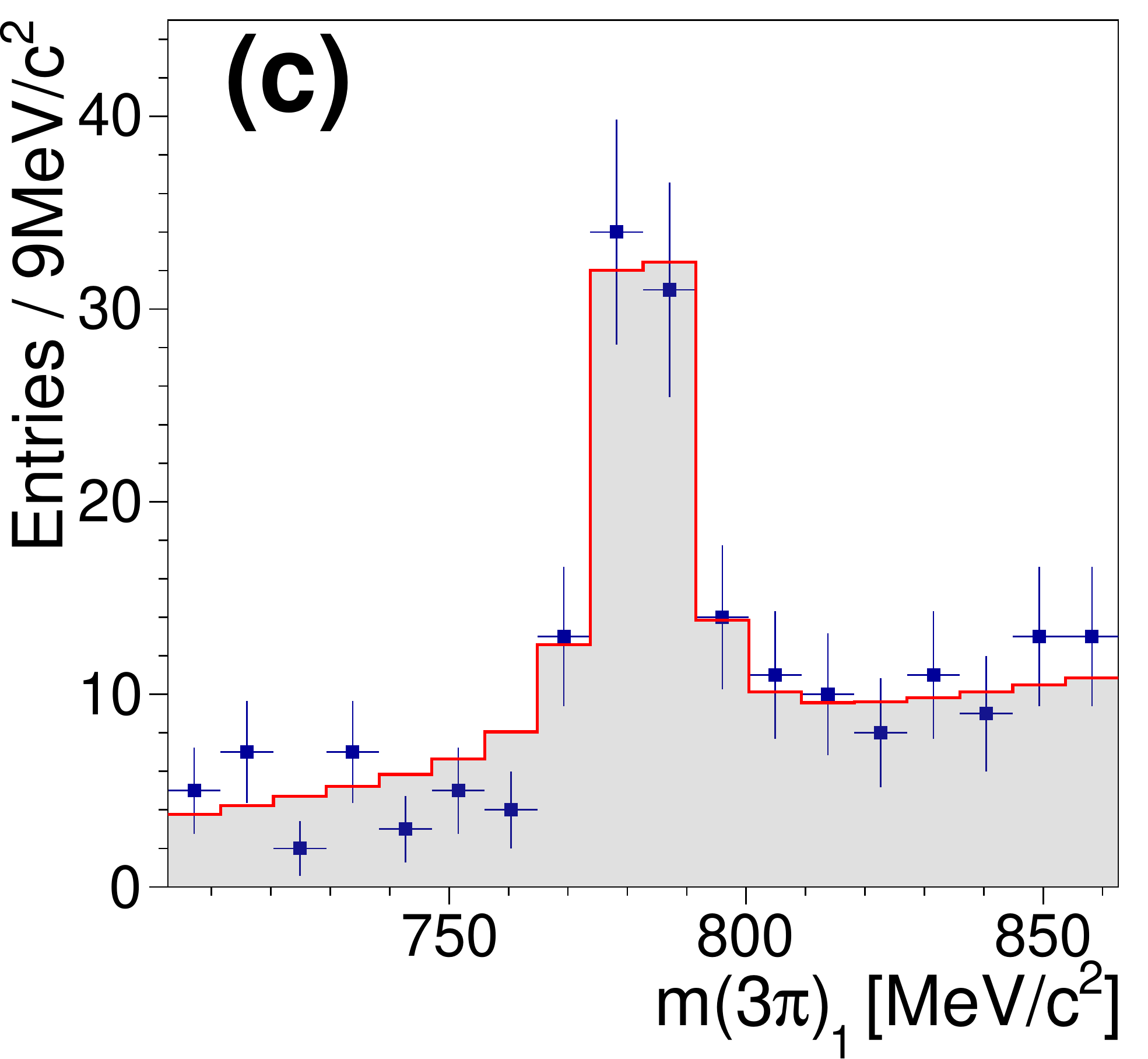} \,\, 
      \includegraphics[width=.22\textwidth]{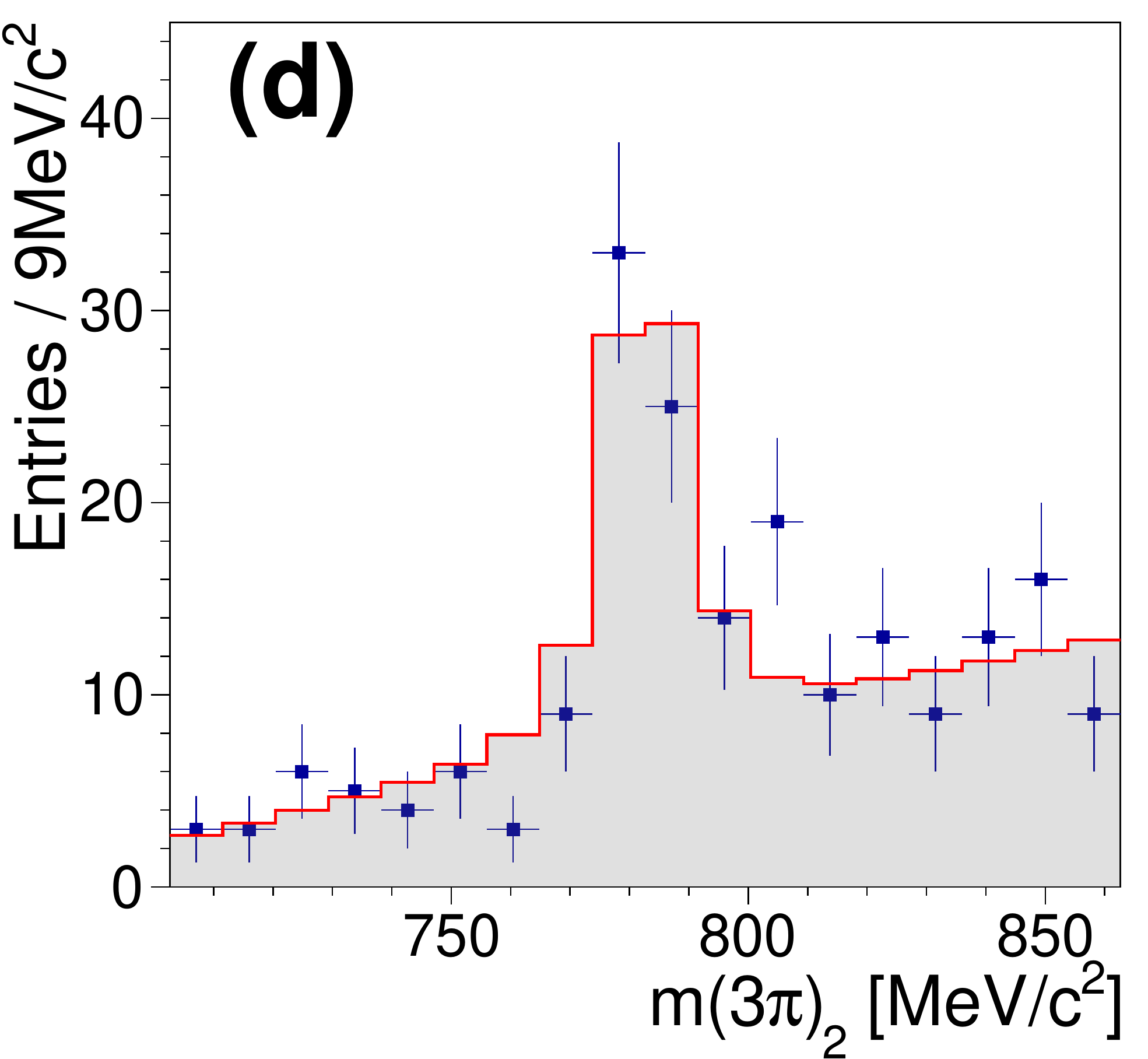}
      \caption{Example of a fit to a data subset of 200 nearest neighbors to a single $\gamma\omega\omega$ event. (a) and (b) show the 3$\pi$ versus $3\pi$ invariant mass distributions for data and the fit function, respectively. For better comparability, (c) and (d) show the projections of the data and fit function to both of the 3$\pi$-axes. }
      \label{fig02}
    \end{figure}

   For this analysis the two-dimensional $m(3\pi)_1$ versus $m(3\pi)_2$ distribution was chosen as the distinct kinematic variable. The signal is described with a two-dimensional Voigtian function, which is defined as the convolution of a Gaussian with a Breit-Wigner function, while the background consists of two different contributions: A two-dimensional linear function with individual slope parameters for the two 3-pion invariant masses is used to describe the homogeneous background.
    Additionally, the $\omega$ bands are described with a Voigtian function for the one, and a linear function for the corresponding other $3\pi$ invariant mass. These functional dependencies are determined using signal MC samples. 
    Figure \ref{fig02}\,(a) shows the $3\pi$ versus $3\pi$ distribution for the $N=200$ nearest neighboring events of a seed event, while Fig. \ref{fig02}\,(b) shows the function fitted to this data. The value of $N$ should be as small as possible to ensure that the phase space cell of all selected neighbors is small and the assumption that the background behaves smoothly within the cell is satisfied, yet it has to be large enough to ensure stable and reliable single-event fits. The value is determined based on dedicated MC studies for this analysis by increasing $N$ until stable fits are achieved. The MC samples are generated using an amplitude model obtained from a PWA fit so that all angular and invariant mass distributions of the recorded data are reproduced.
    The signal-to-background ratio at the location of the seed event is extracted from each single-event fit and represents the $Q$-factor for this event. To illustrate the quality of these fits, the projections of fit function and data from Fig. \ref{fig02} (a) to each of the $3\pi$ axes is shown in the sub-figures (c) and (d), where a good agreement can be seen.

\begin{figure}[t]
      \includegraphics[width=.24\textwidth]{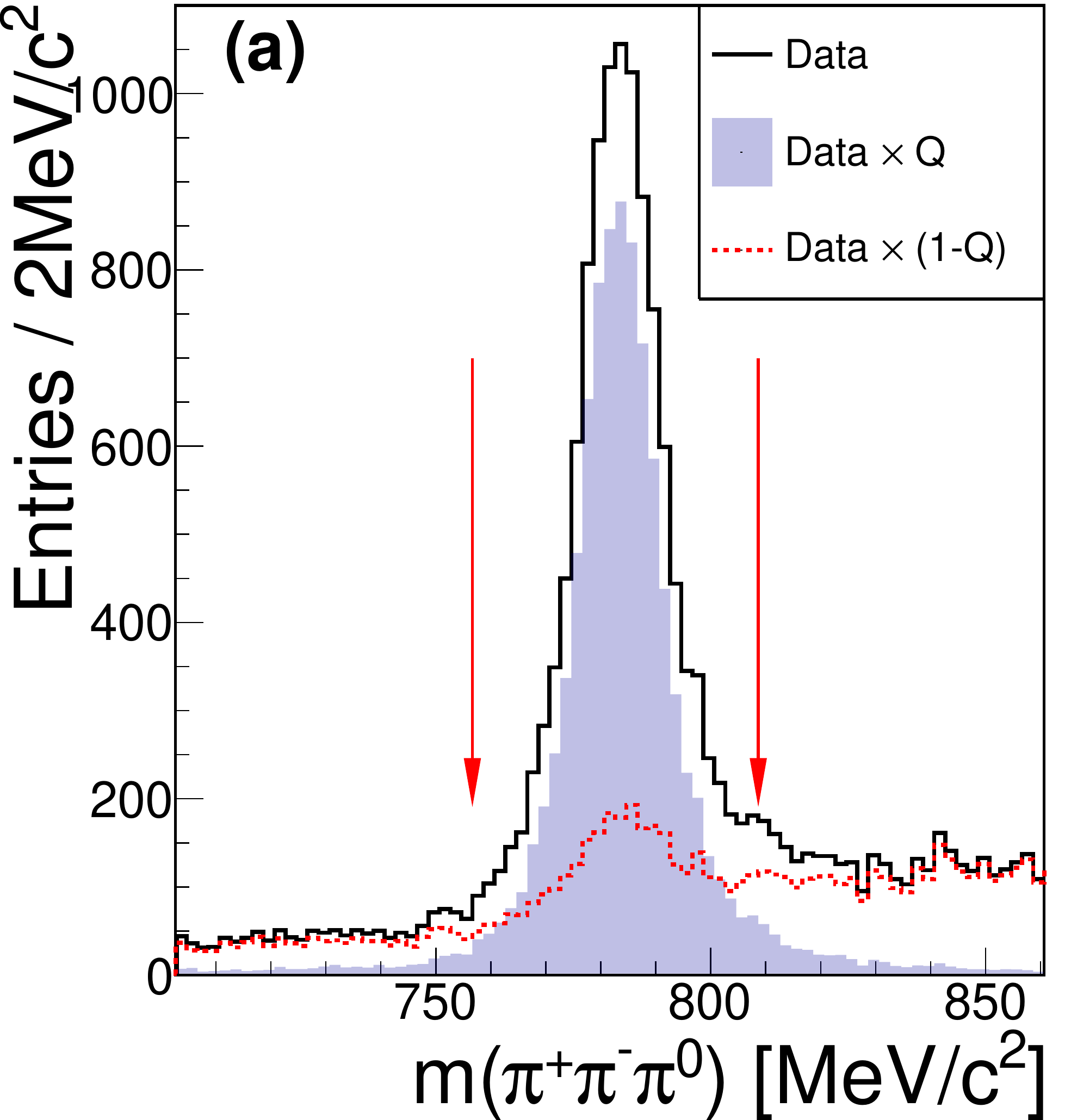}
      \includegraphics[width=.23\textwidth]{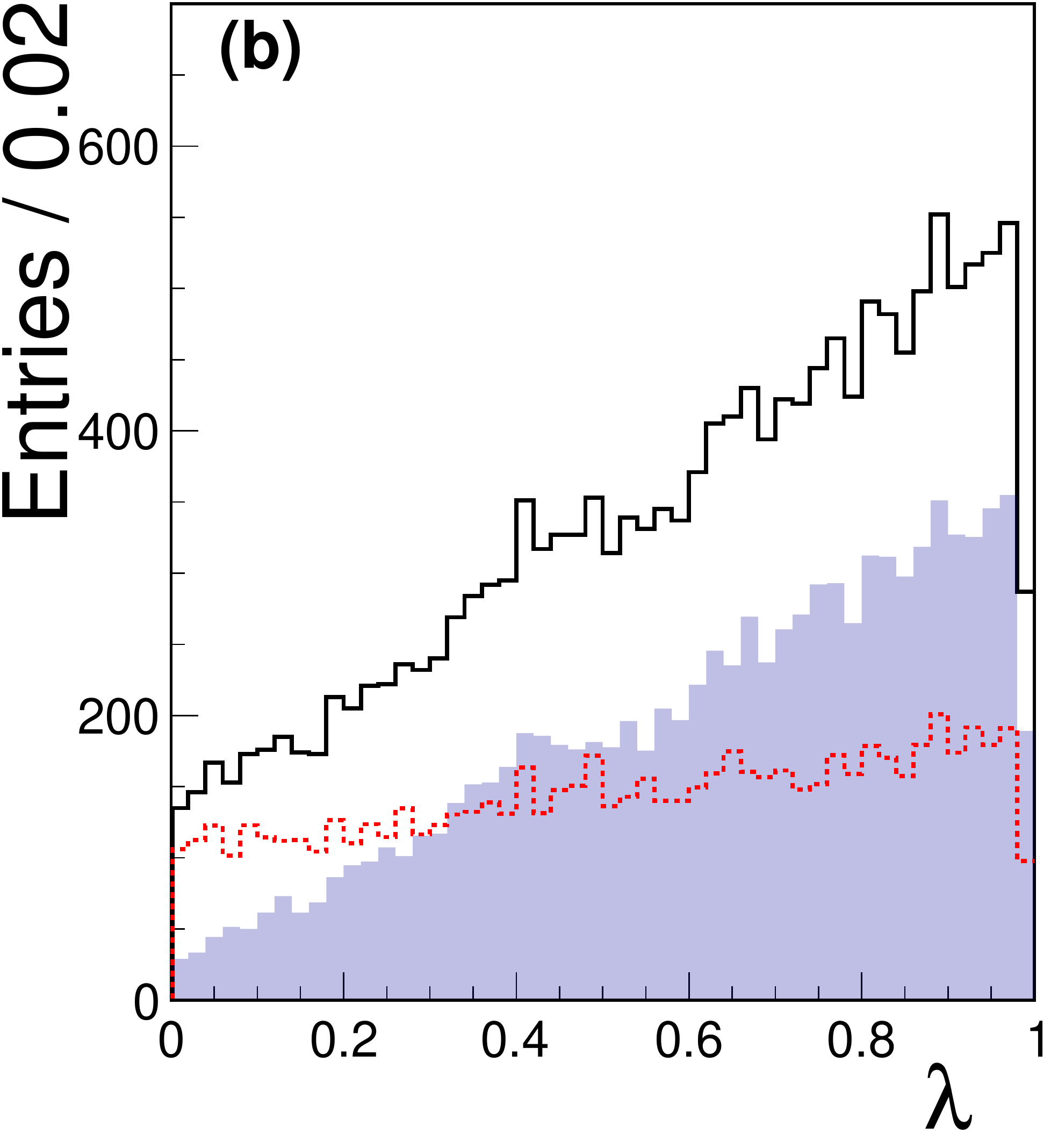}
      \caption{(a) 3$\pi$ invariant mass for all pre-selected events (black), as well as a $Q$-weighted (blue shaded area) and a $(1-Q)$-weighted (red dashed) version of the same distribution. The red arrows indicate the signal region, which is selected after application of the $Q$-factor method. (b) Normalized $\tilde{\lambda}$ distribution for all (black), $Q$-weighted (blue shaded) and $(1-Q)$-weighted (red dashed) events. Both diagrams contain two entries per event, one for each $\omega$ candidate.}
      \label{fig03}
    \end{figure}

    Figure \ref{fig03} shows the invariant $3\pi$ mass and the normalized $\tilde{\lambda}$ distribution for all pre-selected events, as well as the distributions weighted by $Q$ and $(1-Q)$ (both diagrams contain two entries per event, one for each $\omega$ candidate). 
    The $Q$-weighted diagrams show a background-free $\omega$ signal and a linearly increasing $\tilde{\lambda}$ distribution, starting at the origin, as it is expected for a pure $\omega$ signal. 
    
\begin{figure}[t]
      \includegraphics[width=.46\textwidth]{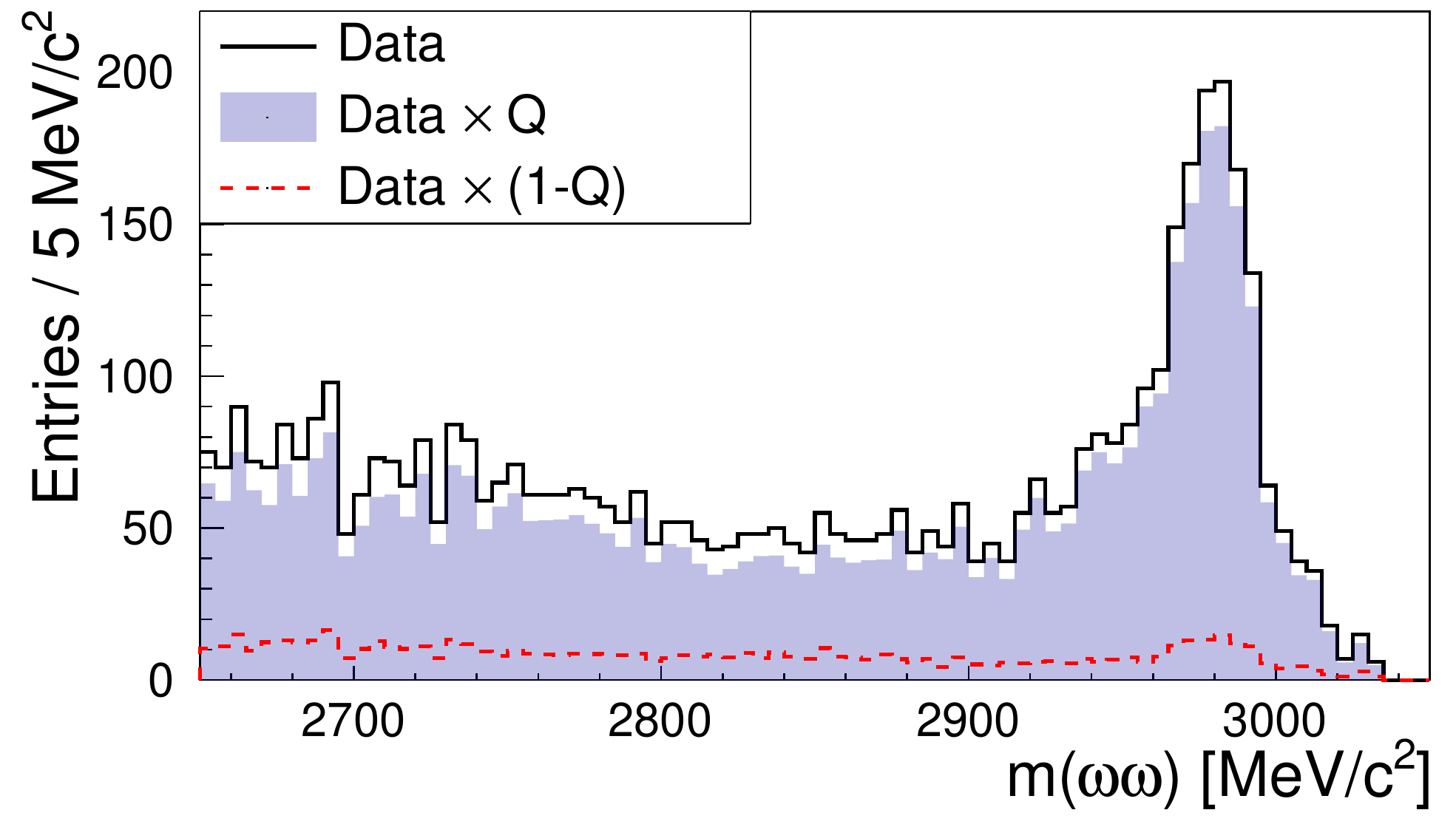} 
      \caption{Invariant $\omega\omega$ mass for selected events, where both $\omega$ candidates lie within a distance of 26\,MeV/$c^2$ from the nominal $\omega$ mass (indicated by the red circle in Fig. \ref{fig01}). The black histogram shows all events in this region, while the blue-shaded area shows the $Q$-weighted and the red-dashed line the $(1-Q)$-weighted version of this distribution, respectively.}
      \label{fig04}
    \end{figure}

    The $(1-Q)$-weighted distributions contain background due to events without any intermediate $\omega$ resonances (linear shape in 3$\pi$ invariant mass, flat distribution of $\tilde{\lambda}$), as well as events that only contain one instead of two $\omega$ mesons. The latter create a peaking structure in the invariant $3\pi$ mass as well as a slight increase of the $(1-Q)$-weighted $\tilde{\lambda}$ distribution. 
    After all single-event fits are performed, the initially very large mass window for the $\omega$ candidates, which is needed to be able to fit the background component underneath the $\omega\omega$ signal, is replaced with a tighter requirement of $26\,$MeV around the two nominal $\omega$ masses, as indicated by the red circle in Fig.~\ref{fig01}. Figure \ref{fig04} shows the invariant $\omega\omega$ mass for the finally selected events within this narrow signal region without any weight, $Q$-weighted and $(1-Q)$-weighted. 
    
    In total 5128 events are selected in the signal region defined as $m(\omega\omega)\geq2.65\,$GeV/$c^2$ and with all other selection criteria applied as discussed above. The sum of the obtained $Q$-factors for these events yields 4489.31, so that about 12.5\% of the initially selected events originate from background sources and are weighted out by the $Q$-factor method. All further analysis steps are performed using this weighted data sample. A strong signal of the $\eta_c$ is observed in this mass distribution.
    
    The performance of the background suppression method is checked by selecting events from side-band regions in the $3\pi$ versus $3\pi$ mass distribution. A very good agreement between expectations from the side bands and the $(1-Q)$-weighted data is found. This underlines the applicability of the method.
    Additionally, as a cross-check and for tuning parameters like the number of neighbors, input-output checks are performed using different MC samples generated with amplitude models obtained from rough fits to the signal and sideband regions. Using the $Q$-factor method, the generated signal and background samples can be identified clearly and the remaining deviation from the generated sample is taken as a systematic uncertainty of the method.

\section{DATA ANALYSIS}
We use a PWA to determine the number of $\eta_c$ candidates and the selection efficiency respecting all dimensions of the phase space simultaneously for the reaction under investigation. The amplitudes are constructed in our software \cite{leapKopf} using the helicity formalism by describing the complete decay chain from the initial $J/\psi$ state to the final state pions and photons. 
We assume that there are no other resonances nearby and thus the selected $\gamma\omega\omega$ events are described either as originating from the decay of the $\eta_c$, or as phase space-like contributions with different $J^P$ quantum numbers of the $\omega\omega$ system, to consider tails of resonances that are located far away from the region of interest. 
For the amplitudes that describe the radiative decay of the $J/\psi$, an expansion into the electromagnetic multipoles of the radiative photon is applied. The decay of the $\eta_c$ as well as the phase space-like contributions are described using an expansion of the corresponding helicity amplitudes into the $LS$-scheme, where $L$ denotes the orbital angular momentum between the two decay products and $S$ their total spin.

\subsection{Amplitude Model}
The differential cross section of the reaction under study is expressed in terms of the transition amplitudes for the production and decay of all intermediate states and is given as  
\begin{align}
          \label{form-Xsec}
          \frac{d\sigma}{d\Omega} \propto w &= \sum_{\lambda_\gamma,M=-1,1} \left| \sum_X \left[ \sum_{\lambda_X} T_{\lambda_\gamma \lambda_X}^{1 M} (J/\psi\rightarrow\gamma X) \right.\right.\nonumber \\ 
                                            &\cdot \sum_{\lambda_{\omega_1}\lambda_{\omega2}} \tilde{A}_{\lambda_{\omega_1}\lambda_{\omega_2}}^{J_X \lambda_X}(X\rightarrow\omega_1\omega_2)\nonumber \\ 
                                            &\left.\left. \cdot A^{J_{\omega_1}}_{\lambda_{\omega_1}} (\omega_1\rightarrow\pi^+_1\pi^-_1\pi^0_1)\cdot A^{J_{\omega_2}}_{\lambda_{\omega_2}} (\omega_2\rightarrow\pi^+_2\pi^-_2\pi^0_2) \right]\right|^2.
        \end{align} 
        Here, $d\Omega$ denotes an infinitesimally small element of the phase-space, and the function $w$ is the transition probability from the initial to the final state.
      The outer (incoherent) sum runs over the helicity of the radiative photon, $\lambda_\gamma$, as well as the $z$-component of the spin of the $J/\psi$, denoted with $M$. Furthermore, for all intermediate states $X$, a coherent summation over the helicity of the state ($\lambda_X$) as well as its daughter particles ($\lambda_{\omega_1},\lambda_{\omega_2}$) is performed. In this expression, $X$ denotes the phase space-like contributions with spin-parity $J^P$, as well as the resonant $\eta_c$ component.
      The amplitudes for the $J/\psi\to\gamma X$ process are given by 
      \begin{align}
         T^{1M}_{\lambda_\gamma \lambda_X} = \sqrt{\frac{3}{4\pi}}d^1_{M\ (\lambda_\gamma -\lambda_X)} (\vartheta) \cdot F^1_{\lambda_\gamma \lambda_X},
         \label{eqn-prodAmp}
       \end{align}
       \noindent where $d$ denotes the Wigner $d$-matrices as defined in Ref.\,\cite{pdg2016}. The $d$-matrices do not depend on the azimuthal angle $\varphi$ in contrast to the usual Wigner $D$-matrices. The $\varphi$ dependence vanishes for the $J/\psi$ decay amplitudes, since both the electron and the positron beams are unpolarized. $F$ represents the complex helicity amplitude, which is then expanded into radiative multipoles related to the corresponding final state photon using the transformation 
     \begin{align}
          \label{form-transfRadMult}
          F_{\lambda_\gamma \lambda_X}^{1} = &\sum_{J_\gamma} \sqrt{\frac{2J_\gamma+1}{3}} \cdot \frac{B_{L_\text{min}}(q)}{B_{L_\text{min}}(q_0)}\nonumber\\
                                                      &\cdot\langle J_\gamma,\lambda_\gamma;1,\lambda_X-\lambda_\gamma|J_X,\lambda_X\rangle a_{J_\gamma},
        \end{align}
        \noindent as given in Refs.\,\cite{HeliMultipole1} \cite{HeliMultipole2} \cite{HeliMultipole3}, where $\langle ... \rangle$ denotes the Clebsch-Gordan coefficients and $B_L(q)$ are the Blatt-Weisskopf barrier factors as defined in Ref.\,\cite{Chung:1993da}. Here, $q$ is the linear momentum of one of the decay products in the $J/\psi$ rest frame. $q_0$ is chosen as the breakup momentum for the $X$ system and to coincide with the $\omega\omega$ mass threshold. Since the orbital angular momentum $L$ between the decay products is not defined in the multipole basis, we use the minimal value $L_\text{min}$ depending on the spin-parity of $X$, which is expected to represent the dominant contribution.
     Due to this transformation, the helicities are replaced by a description based on the angular momentum $J_\gamma$ carried by the radiative photon. This way, the single terms of the expansion can be identified with electric or magnetic dipole, quadrupole and octupole transitions.

     The decay amplitudes $\tilde{A}$ are given by
     \begin{align}
        \tilde{A}_{\lambda_{\omega_1}\lambda_{\omega_2}}^{J_X \lambda_X} = \sqrt{\frac{2J_X+1}{4\pi}} D^{J_X *}_{\lambda_X\ (\lambda_{\omega_1}-\lambda_{\omega_2})}(\varphi,\vartheta,0) \cdot F^{J_X}_{\lambda_{\omega_1}\lambda_{\omega_2}}.
        \label{eqn-decAmpl}
     \end{align}
     \noindent For these amplitudes an expansion into states with defined sets of $J^{PC}$, $L$, $S$ values is performed using the transformation 
      \begin{align}
          \label{form-transfLS}
          F_{\lambda_{\omega_1}\lambda_{\omega_2}}^{J_X} &= \sum_{L,S} \sqrt{\frac{2L+1}{2J_X+1}} \cdot \frac{B_{L}(q)}{B_{L}(q_0)} \nonumber \\
                      &\cdot \langle L,0;S,\lambda_X|J_X,(\lambda_{\omega_1}-\lambda_{\omega_2})\rangle \nonumber \\
                      &\cdot \langle s_{\omega_1},\lambda_{\omega_1};s_{\omega_2},-\lambda_{\omega_2}|S,\lambda_X\rangle\cdot\alpha_{LS}^{J_X},
        \end{align}
        \noindent where $S$ is the total spin of the $\omega\omega$ system \cite{chung}. Also here, the normalized Blatt-Weisskopf factors are included as defined above. For the $\eta_c$ component, the break-up momentum $q_0$ is chosen to coincide with the nominal mass of the $\eta_c$, while for all other contributions the $\omega\omega$ mass threshold is used. Since we assume that no resonances apart from the $\eta_c$ are nearby, the description of the dynamical part of the amplitudes for the phase space-like components (\emph{e.g.} Breit-Wigner function) is omitted. For the line shape of the $\eta_c$ a modified relativistic Breit-Wigner function is used that takes the distortion due to the pure magnetic dipole transition $J/\psi\to\gamma\eta_c$ into account. The amplitude is modified by a factor $E_\gamma^{3/2}$, which originates from the $M1$-transition matrix element \cite{brambilla} and corresponds to the expected $E_\gamma^3$ dependency of the observed line shape. Since this factor leads to a good description around the pole mass but also introduces a diverging tail towards larger energies of the radiative photon (smaller invariant $\omega\omega$ masses), the amplitude for the $\eta_c$ is further modified using an empirical damping factor $\exp\left(-\frac{E_\gamma^2}{16\beta^2}\right)$ with $\beta=0.065\,$GeV, in accordance with the factor used by the CLEO collaboration \cite{cleo-etac-erratum}. 

     The decay amplitudes $A$ of the $\omega$ resonances are directly proportional to the parameter $\tilde{\lambda}$ introduced in Eq.\,(\ref{eqn2}).
        The normal vector $\vec{n}$ to the $\omega$ decay plane spanned by the three daughter particles in its helicity frame is described in terms of the Euler angles $\vartheta_n$, $\varphi_n$ and $\gamma_n=0$. With $\mu = \vec{J}_\omega\cdot\vec{n}$ being the projection of the $\omega$ mesons spin to the direction of $\vec{n}$, the amplitude reads as
        \begin{align}
            A^{J_{\omega}}_{\lambda_\omega} (\omega\rightarrow\pi^+\pi^-\pi^0) = \sqrt{\frac{3}{4\pi}}\cdot D_{\lambda_\omega\mu}^{1*}(\varphi_n,\vartheta_n,0)\cdot\tilde{\lambda}_\mu,
            \label{eqn-omDecay}
        \end{align}
     where only the case $\mu=0$ is allowed for this decay \cite{Amsler:2014xta}.

     The free parameters varied in the minimization are the complex values $a_{J_\gamma}$ and $\alpha_{LS}^{J_X}$, as well as the mass and width of the $\eta_c$. Symmetries arising from parity conservation and the appearance of two identical particles ($\omega\omega$) are respected and lead to a reduction of free parameters in the fit.

\subsection{Fit procedure}
Unbinned maximum likelihood fits are performed for all hypotheses, in which the probability function $w$ is fitted to the selected data by varying the free parameters given by the complex amplitudes as well as the masses and widths, if applicable. Each amplitude can be expressed by a real magnitude and a phase, yielding two distinct fit parameters per amplitude.
      The likelihood function is given by \cite{Amsler:2014xta}
      \begin{align}
        \label{form-LH}
        \mathcal{L} \propto N!\cdot\exp\left(-\frac{(N-\overline{n})^2}{2N}\right) \prod_{i=1}^{N}\frac{w(\vec{\Omega}_i,\vec{\alpha})}{\int w(\vec{\Omega},\vec{\alpha})\epsilon(\vec{\Omega})d\Omega},
      \end{align}
        where $N$ denotes the number of data events, $\overline{n}$ is defined as   
        \begin{align}
        \overline{n} = N\cdot\frac{\int w(\vec{\Omega},\vec{\alpha})\epsilon(\vec{\Omega})d\Omega}{\int \epsilon(\vec{\Omega})d\Omega},
       \end{align}
     \noindent $\vec{\Omega}$ is a vector of the phase-space coordinates and $\vec{\alpha}$ of the complex fit parameters. The function $w(\vec{\Omega},\vec{\alpha})$ is the transition probability function given in Eq.\,(\ref{form-Xsec}) and $\epsilon(\vec{\Omega})$ is the acceptance and reconstruction efficiency at the position $\vec{\Omega}$. 

       The function $w$ is interpreted as a probability density function and the corresponding probabilities for all events are multiplied to obtain the total probability.
       A normalization of the extended likelihood function is achieved due to the exponential term in which $\overline{n}$ appears, so that the mean weight of an MC event is approximately $1$ after the likelihood has been maximized.
       The integrals appearing in the $\overline{n}$ term as well as the denominator in the product in Eq.\,(\ref{form-LH}) are approximated using reconstructed, phase space distributed MC events. The events of the MC sample are propagated through the BESIII detector, reconstructed and selected with the same cuts as the data sample to account for the geometrical acceptance and selection efficiency in all dimensions of the phase-space. 
       
       The best description of the data sample is reached upon maximization of the likelihood $\mathcal{L}$. Equation\,(\ref{form-LH}) is logarithmized so that the product is transformed into a sum. Finally, the event weights $Q_i$ obtained from the $Q$-factor method are also included in the likelihood function and a negative sign is added to the logarithmized function, so that commonly used minimizers and algorithms, in this case \textsc{Minuit2} \cite{James:1975dr}, can be used.

       The negative log-likelihood function, which is actually minimized, now reads as
       \begin{align}
          -\ln \mathcal{L} = &-\sum_{i=1}^{N}\ln(w(\vec{\Omega_i},\vec{\alpha}))\cdot Q_i \nonumber\\ 
          &+ \left(\sum_{i=1}^{N}Q_i\right)\cdot \ln\left(\frac{\sum_{j=1}^{n_\textnormal{MC}}w(\vec{\Omega_j},\vec{\alpha})}{n_\textnormal{MC}}\right) \nonumber \\ 
          &+ \frac{1}{2}\cdot \left(\sum_{i=1}^{N}Q_i\right)\cdot \left(\frac{\sum_{j=1}^{n_\textnormal{MC}}w(\vec{\Omega_j},\vec{\alpha})}{n_\textnormal{MC}}-1\right)^2.
       \end{align}

\subsection{Fit strategy}
      \begin{table*}
        \caption{Results of PWA fits for the best five hypotheses.}
        \small
        \begin{center}
           \begin{tabular}{p{1.5cm}p{3cm}p{1.5cm}p{3cm}p{2cm}p{2cm}} 
        \hline
        \hline
        $\mathbf{i}$  & \textbf{Hypothesis}                 &   $\mathbf{-\ln(\mathcal{L})}$  & \textbf{number of}               & $\textit{\textbf{BIC}}$               & $\textit{\textbf{AIC}}$          \\
                      & $\mathcal{H}_i$                     &                                 & \textbf{free parameters}         &                              &                           \\
        \hline
        0  & $\eta_c, 0^{-}, 1^{+}, 2^{+}$                &$-4150.44$       & \hspace{1cm}$21$               &  $-8124.28$     &    $-8258.88$   \\
        1  & $\eta_c, 0^{-}, 2^{+}$                       &$-4130.97$       & \hspace{1cm}$17$               &  $-8118.98$     &    $-8227.94$   \\
        2  & $\eta_c, 0^{-}, 0^{+}, 2^{+}$                &$-4130.93$       & \hspace{1cm}$21$               &  $-8085.26$     &    $-8219.86$   \\
        3  & $\eta_c, 0^{-}, 0^{+}, 1^{+}$                &$-4113.13$       & \hspace{1cm}$13$               &  $-8116.95$     &    $-8200.27$   \\
        4  & $\eta_c, 0^{-}, 0^{+}$                       &$-4058.43$       & \hspace{1cm}$9$                &  $-8041.17$     &    $-8098.85$   \\
        \hline
        \hline
        \end{tabular}
        \end{center}
        \label{tab-etacResults}
      \end{table*}
Since the composition of the non-resonant contribution is not known a priori, different hypotheses are fitted to the selected data set, which contain the $\eta_c$ component and one up to a maximum of four different non-resonant components. These non-resonant components are assumed to have the $J^P$ quantum numbers $0^-$, $0^+$, $1^+$ or $2^+$, so that the most simple hypothesis is given as $\{\eta_c, 0^-\}$, and the most complex one by $\{\eta_c, 0^-, 0^+, 1^+, 2^+\}$.
We also perform fits including higher spin contributions ($J^P=4^+$) and the contribution of a spin-4 component is found to be not significant. Similarly, fits with contributions carrying exotic quantum numbers (\emph{e.g.} $J^{PC}=1^{-+}$) as well as pseudo tensor contributions ($J^{PC}=2^{-+}$) are tested and found to be insignificant.
        
        In order to be able to compare the quality of fits with different, generally not nested, hypotheses with different numbers of free parameters, two information criteria from model selection theory are utilized.
        The Bayesian Information Criterion ($BIC$) depends on the maximized value of the likelihood $\mathcal{L}$, the number of free parameters $k$ as well as the number of data points $n$, which is given by the sum of the $Q$-factors. It is defined as
        \begin{align}
          \label{form-BIC}
          BIC = -2\cdot \ln(\mathcal{L}) + k\cdot \ln(n).
        \end{align}
      The $BIC$ is based on the assumption that the number of data points $n$ is much larger than the number of free parameters $k$~\cite{modelSelection}. This assumption is fulfilled for all fits performed here. 

      The second criterion is the Akaike Information Criterion ($AIC$), which provides a different penalty factor compared to the $BIC$. 
      It is defined as  
      \begin{align}
        \label{form-AICc}
        AIC &=  -2\cdot \ln(\mathcal{L}) + 2\cdot k, 
      \end{align}
      
      \noindent thus it is independent from the sample size $n$. In comparison to the $BIC$, the penalty term is much weaker, which increases the probability of over-fitting. 
       
      \noindent Theoretical considerations show \cite{modelSelection} that in general $AIC$ should be preferred over $BIC$ due to reasons of accurateness as well as practical performance. 

\begin{figure}[ht]
   \centering
   \includegraphics[width=.48\textwidth]{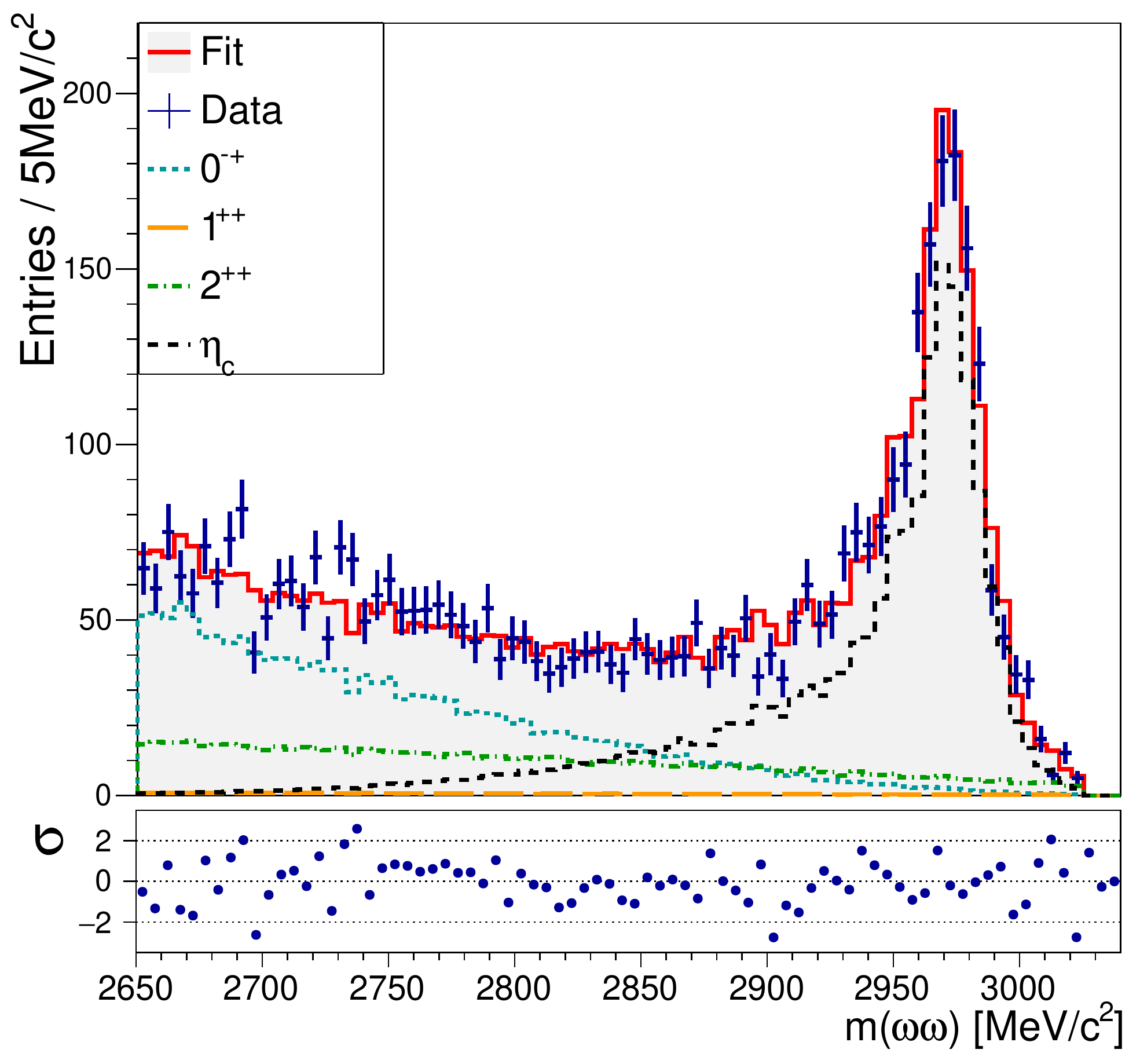} \\ 
   \includegraphics[width=.48\textwidth]{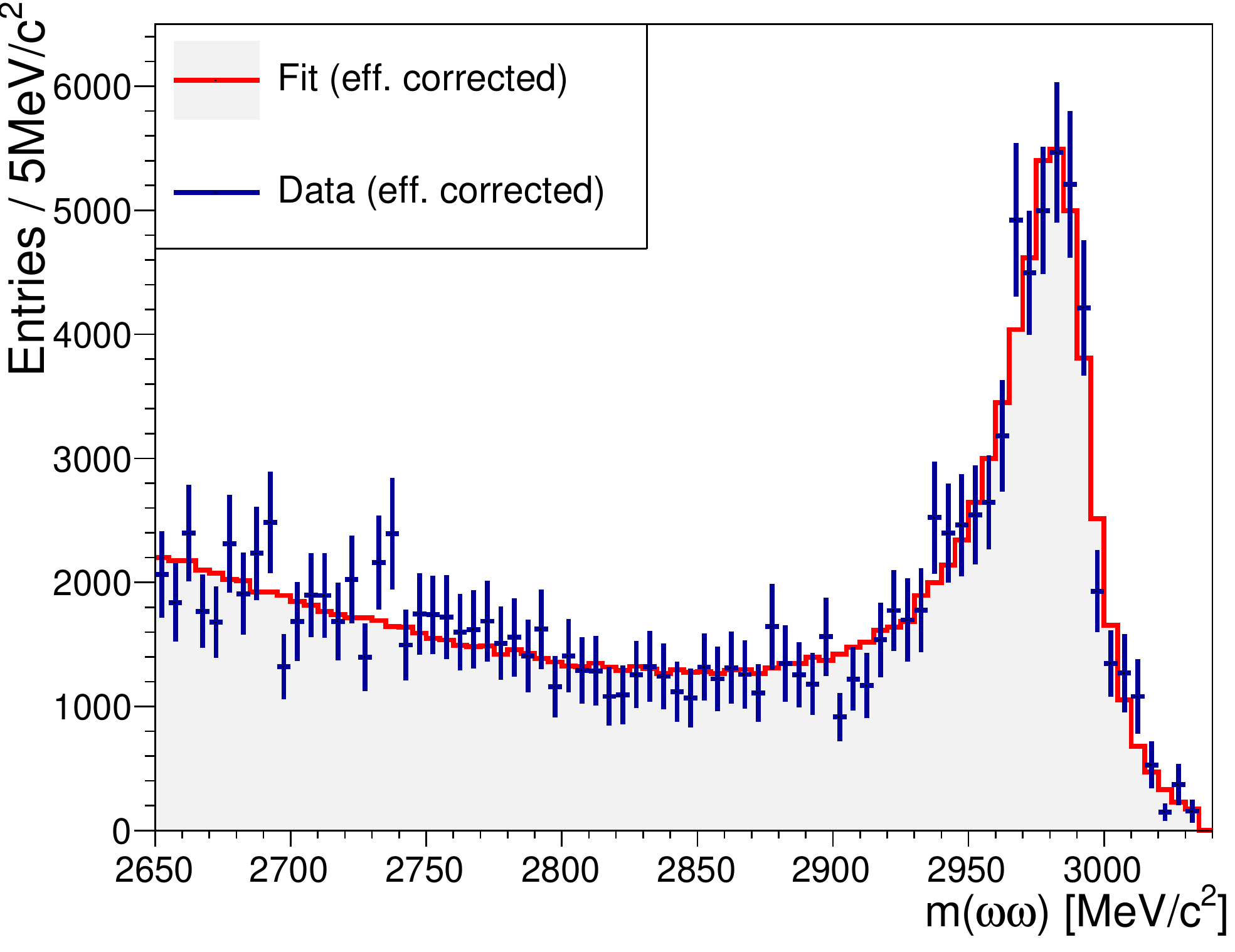} 
   \caption{Projection of the best fit and its individual components to the invariant $\omega\omega$ mass. The residuals are shown below the mass spectrum in units of the statistical error. The lower plot shows an efficiency and acceptance corrected version of the same invariant mass spectrum.}
   \label{fig05}
\end{figure}

\begin{figure*}
     \centering
     \includegraphics[width=.95\textwidth]{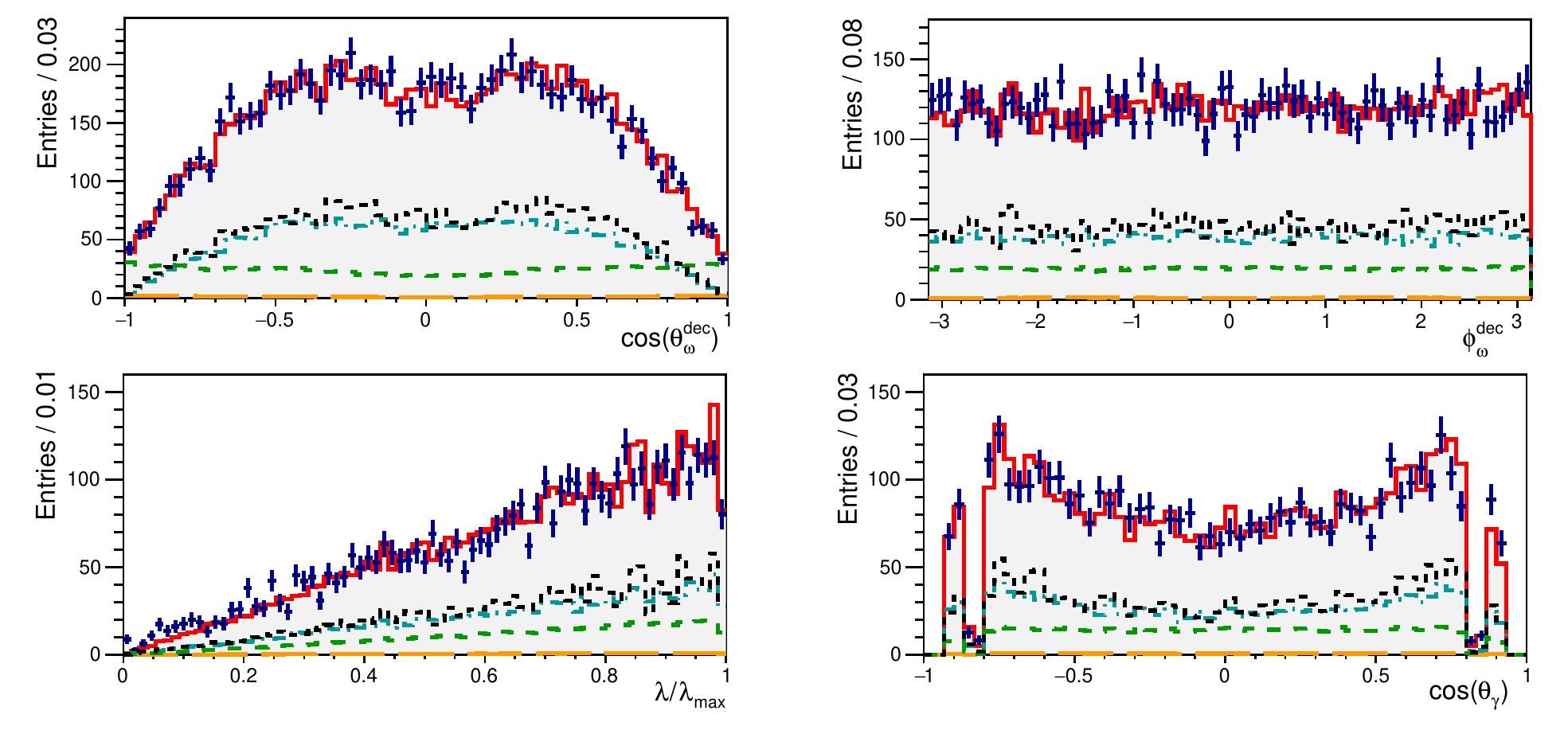} \\
     \includegraphics[width=.95\textwidth]{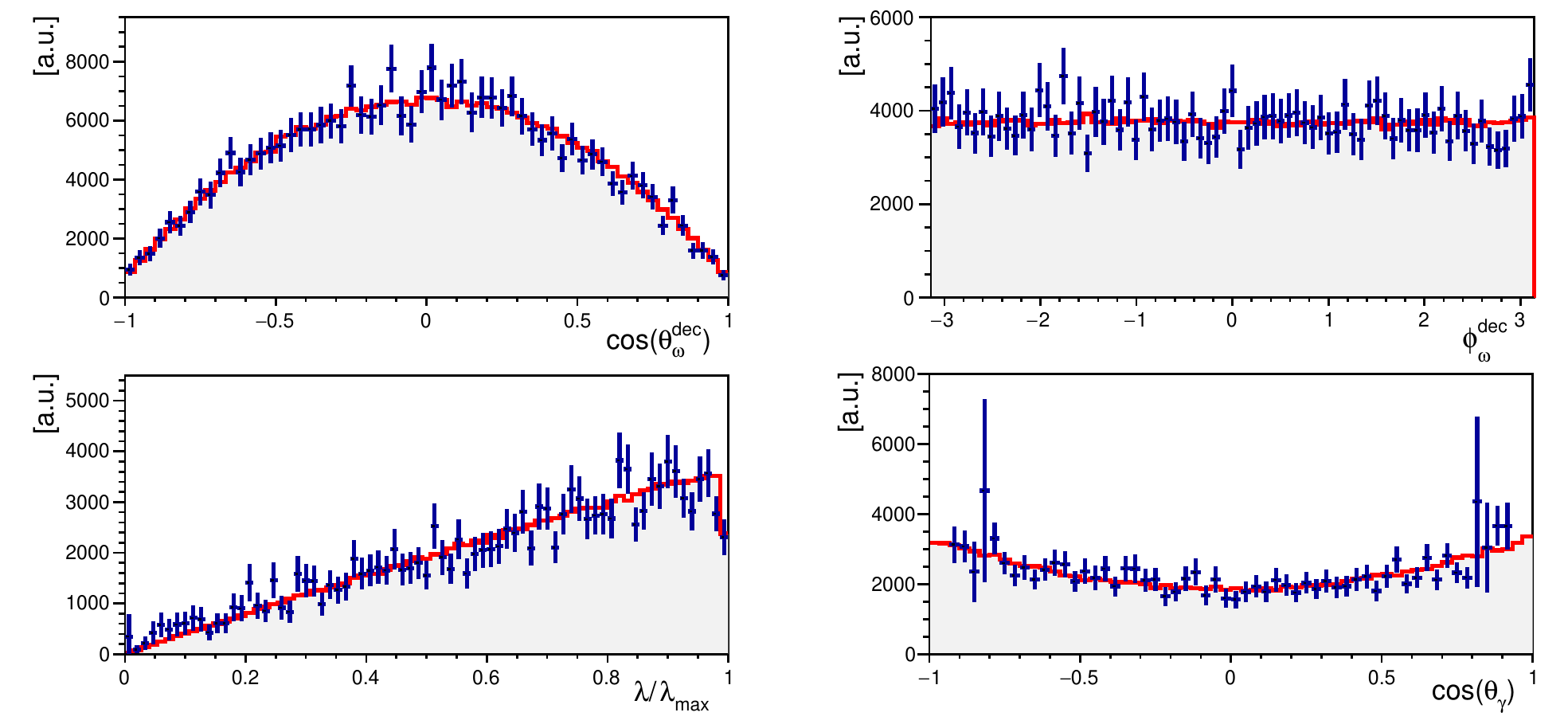}
     \caption{Projections of the best fit and the individual fit components to the polar (upper left) and azimuthal (upper right) decay angle of the $\omega$ mesons in the corresponding $\omega$ helicity frame, the normalized $\tilde{\lambda}$ distribution and the polar angle of the radiative photon in the $J/\psi$ helicity frame. The lower two rows show the efficiency and acceptance corrected versions of the plots described above. The same color and line-style code as in Fig. \ref{fig05} is used.}
  \label{fig06}
\end{figure*}

      As for the likelihood, also for $BIC$ and $AIC$ a more negative value indicates a better fit. The results for the five best hypotheses are listed in Table \ref{tab-etacResults}. 
      The overall best hypothesis is determined to be
      \begin{align}
         \mathcal{H}_0 = \{\eta_c, 0^-, 1^+, 2^+\},
      \end{align}
      for which 21 parameters are free in the fit. The fit parameters are composed of the complex decay amplitudes $a_{J_\gamma}$ for the process $J/\psi\to\gamma X$ in the radiative multipole schema, as well as the $X\to\omega\omega$ decay amplitudes $\alpha^{J_X}_{LS}$ after the transformation to the $LS$-scheme and the mass and width of the $\eta_c$. Each complex decay amplitude yields two independent fit parameters (magnitude and phase), whereas the phase parameter for the $J/\psi\to\gamma \eta_c$ amplitude is fixed to zero as a global reference. Additionally, one magnitude and one phase parameter are fixed for the $X\to\omega\omega$ decay amplitudes for each of the four fit contributions to obtain a set of independent parameters.
      A projection of this fit to the $\omega\omega$ invariant mass and other kinematically relevant variables is shown in Figs. \ref{fig05} and \ref{fig06}. These figures also show efficiency-corrected versions of all mass spectra and angular distributions. The correction is performed using the PWA software and is therefore done in all dimensions of the phase-space simultaneously. The fit yields a total of $1705\pm58$ $\eta_c$ events, which is the number used for the calculation of the BF. The yields of all components are listed in Table \ref{tab-yields}.
      \begin{table}
         \caption{Yields and fit fractions of single components for the best fit. The fit fraction is defined as the ratio of the intensity of a single component to the total intensity. The sum of all single components sums up to only 87.6\% due to interference effects.}
         \begin{tabular}{lll}
            \hline
            \hline
            \textbf{Component\ \ \ \ \ \ \ \ }    & \textbf{Yield\ \ \ \ \ \ \ \ \ \ }& \textbf{Fit Fraction}  \\
            \hline
              $0^{-}$            & $1462\pm95$ & $(32.6\pm2.2)\%$\\
              $1^{+}$            & $37\pm20$ & $(0.8\pm0.4)\%$\\
              $2^{+}$            & $727\pm89$ & $(16.2\pm2.0)\%$\\
              $\eta_c$           & $1705\pm58$ & $(38.0\pm2.1)\%$\\
            \hline
         \end{tabular}
         \label{tab-yields}
      \end{table}

      
     To estimate the overall goodness-of-fit, a global $\chi^2$ value is calculated by comparing the histograms for data and fit projections in all relevant kinematic variables as defined for the metric used for the $Q$-factor background subtraction method (see Section \ref{williams}). The global reduced $\chi^2$ is calculated as
      \begin{align}
          \frac{\chi^2}{ndf} = \sum_i \sum_{j=0}^{N_{\text{bins},i}} \frac{(N^\text{data}_{ij}-N^\text{fit}_{ij})^2}{(\sigma^\text{data}_{ij})^2+(\sigma^\text{fit}_{ij})^2} / (N_\text{bins}-N_\text{params}),
      \end{align} 
      \noindent where $N^\text{data}_{ij}$ and $N^\text{fit}_{ij}$ are the contents of the $j$th bin in the $i$th kinematic variable for data and fit histograms, respectively. The bin contents themselves are given by the sum of weights of the events for data ($Q$-weights) as well as fit (weights from the PWA fit) histograms. Accordingly, $\sigma^\text{data}_{ij}$ and $\sigma^\text{fit}_{ij}$ represent the corresponding sum of squared weights to account for the bin error in the weighted histograms. $N_\text{bins}$ is the sum of all bins considered and $N_\text{params}$ is the number of free parameters in the PWA fit. Bins with less than 10 effective events are merged with neighboring bins. For the best fit hypothesis $\mathcal{H}_0$, a value of $\chi^2/ndf = 640 / (609-21) = 1.09$ is obtained, which indicates a good quality of the fit.

\section{SYSTEMATIC UNCERTAINTIES}
  Various sources of systematic uncertainties for the determination of the BF, the mass and the width of the $\eta_c$ are considered. The uncertainties arise from the reconstruction and fit procedure, background subtraction method, external BFs, kinematic fit, parameterization of the $\eta_c$ line shape and the number of $J/\psi$ events in our data sample. 
  \paragraph{Number of $J/\psi$ events}
    Inclusive decays of the $J/\psi$ are used to calculate the number of $J/\psi$ events in the data sample used for this analysis. The sample contains $(1310.6\pm7.0)\times10^6$ $J/\psi$ decays, where the uncertainty is systematic only and the statistical uncertainty is negligible \cite{Ablikim:2016fal}. The uncertainty propagates to a systematic uncertainty on the $\eta_c\to\omega\omega$ BF of 0.5\%.
  \paragraph{Photon detection}
    The detection efficiency for photons is studied using the well understood process $J/\psi\to\pi^+\pi^-\pi^0$. A systematic uncertainty introduced by the photon reconstruction efficiency of $<1\%$ per photon is found. The systematic uncertainty for the reconstruction of the five signal photons in this analysis thus is conservatively taken to be 5\%.
  \paragraph{Track reconstruction}
    For the estimation of the systematic uncertainty arising from the reconstruction of charged tracks and the identification of pions, a detailed study of the process $J/\psi\to p\overline{p}\pi^+\pi^-$ is performed. It is found that a systematic uncertainty of 1\% per pion is a reasonable estimation  and thus the corresponding systematic uncertainty for the four charged pions in this analysis is set to 4\%. 
  \paragraph{External branching fractions}
  The uncertainties of the BFs entering this analysis, namely those of the decays $J/\psi\to\gamma\eta_c$ and $\omega\to\pi^+\pi^-\pi^0$ are taken from the world average values published in Ref.\,\cite{pdg2016} and treated as systematic uncertainties. The uncertainty of $\mathcal{B}(\pi^0\to\gamma\gamma)$ is negligible and is therefore excluded from Table \ref{tab-systUncert}. It should be noted here that the uncertainty on the BF $J/\psi\to\gamma\eta_c$ is the dominant uncertainty in this analysis.
  \paragraph{Kinematic fit}
  To estimate the systematic uncertainty of the kinematic fit, the charged track helix parameters in simulated data are smeared with a Gaussian function so that their distributions in MC and data match. The difference in efficiency between applying and not applying this correction for the given requirement on the $\chi^2_{6C}$ value of the kinematic fit is found to be $1.2\%$ and is taken as the systematic uncertainty. 
  \paragraph{$Q$-factor method}
    To estimate the systematic uncertainty introduced by the $Q$-factor method, tests with different dedicated MC samples are performed. Background and signal MC samples of different compositions are generated and subjected to the $Q$-factor method. The largest deviation between the number of generated signal events and the sum of the obtained $Q$-factors is obtained using a background sample that contains a peaking background contribution at the mass of the $\eta_c$ among other phase space-like contributions. The deviation is determined to be 0.9\%, which is taken as the systematic uncertainty of the method.
  
  \begin{table}[t]
     \caption{Summary of all systematic uncertainties listed by their source. If the determination of a systematic uncertainty is not applicable for a given variable, the corresponding field is filled with a dash line.}
          \begin{tabular}{llll}
          \hline
          \hline
             \textbf{Source}                           & $\mathcal{B}$\hspace{20pt}  & $M(\eta_c)$ & $\Gamma(\eta_c)$ \\
                                                    & $(\%)$        & (MeV$/c^2$) & (MeV) \\
          \hline
          Number of $J/\psi$ events                 & $0.5$  & --- & ---\\
          Photon detection                          & $5.0$    & --- & ---\\
          Track reconstruction                      & $4.0$    & --- & ---\\
          External branching fractions:             &          &     &\\
          \ \ \ $J/\psi\rightarrow\gamma\eta_c$     & $23.5$ & --- & ---\\
          \ \ \ $\omega\rightarrow\pi^+\pi^-\pi^0$  & $0.8$  & --- & ---\\
          Kinematic Fit                             & $1.2$ & --- & ---\\
          $Q$-factor method & $0.9$& --- & ---\\
          $\eta_c$ damping factor                   & $14.2$ & $0.3$ & $1.8$\\
          Variation of fit range                    &  $1.4$      & $0.2$ & $0.6$ \\
          $\eta_c$ resonance parameters & $1.0$   & --- & ---\\
          Selection of fit hypothesis                   &  ---      & $0.6$ & $0.3$ \\
          Detector resolution                       &  ---      & $2.0$ & $3.6$ \\
          \hline
             Quadratic Sum                          &           &  \\
             \ \ \ all:                             & $28.3$ & $2.1$ & $4.1$\\
             \ \ \ w/o $\mathcal{B}(J/\psi\rightarrow\gamma\eta_c)$: & $15.8$ & &\\
          \hline
          \hline
          \end{tabular}
          \label{tab-systUncert}
        \end{table}

  \paragraph{$\eta_c$ damping factor}
  To estimate the uncertainty due to the $\eta_c$ damping factor, an alternative parameterization of this factor is used. For this test the CLEO parameterization is exchanged by the function $E_{\gamma,0}^2/(E_\gamma E_{\gamma,0} + (E_\gamma-E_{\gamma,0})^2)$, where $E_\gamma$ denotes the energy of the radiative photon and $E_{\gamma,0}$ is the most probable photon energy, corresponding to the mass of the $\eta_c$ \cite{kedr-etac}. The number of $\eta_c$ events and the efficiency are extracted from this fit and the difference between the resulting BF and the nominal result is measured to be $14.2\%$, which is assigned as a systematic uncertainty. The mass and width of the $\eta_c$ are left floating in this fit and their differences to the nominal result are considered as systematic uncertainties for the measurement of the resonance parameters.
  \paragraph{Fit range} 
    While for the nominal result only events in the region $m(\omega\omega)>2.65\,$GeV$/c^2$ are used, this lower mass limit is varied by $\pm50\,$MeV$/c^2$ to estimate the uncertainty connected to the choice of the mass requirement. The partial wave fit is re-performed for both scenarios and the largest deviation in the yield of the $\eta_c$ candidates is found to be $1.4\,\%$. This value is taken as the systematic uncertainty due to the choice of the fitting mass range. Similarly, also the mass and width of the $\eta_c$ are re-evaluated and the differences to the nominal result are taken as systematic uncertainties.
  \paragraph{$\eta_c$ resonance parameters}
    We also re-performed the fit using fixed values for the resonance parameters of the $\eta_c$. For this study, mass and width are set to their world average values published in Ref.\,\cite{pdg2016} and a deviation of $1.0\,\%$ for the obtained yield of the $\eta_c$ signal is found, which is taken as a systematic uncertainty for the BF discussed in this paper.
  \paragraph{Selection of fit hypothesis}
    The results for the yield, mass and width of the $\eta_c$ are additionally evaluated for the second best hypothesis to estimate the uncertainty due to the choice of the hypothesis. The difference in the obtained number of observed $\eta_c$ events has a negligible effect on the extracted BF. The deviation of the mass is determined to be $0.6\,$MeV$/c^2$ while the width differs by $0.3$ MeV, which are taken as systematic uncertainties.

      \paragraph{Detector resolution}
      To estimate the effect of the detector resolution we perform a dedicated MC study. Using all parameters obtained from the best PWA fit to data, we generate an MC sample and propagate the events through the BESIII detector simulation and reconstruction using the same criteria as for beam data. After performing a PWA fit to the reconstructed and selected MC sample we obtain a difference of $2.0\,$MeV$/c^2$ for the mass and $3.6\,$MeV for the width of the $\eta_c$ between the generated and reconstructed data sample. We use this deviation as an estimation for the systematic uncertainty due to the detector resolution.

\section{BRANCHING FRACTION}
  Using the obtained results of the best fit to the data and the systematic uncertainties discussed above, the product BF of the decay chain $J/\psi\to\gamma\eta_c\to\gamma\omega\omega$ is determined as 
  \begin{align}
     \mathcal{B}&(J/\psi\to\gamma\eta_c)\cdot\mathcal{B}(\eta_c\to\omega\omega) \nonumber \\
     &= \frac{N_{\eta_c}}{N_{J/\psi}\mathcal{B}^2(\omega\to\pi^+\pi^-\pi^0)\mathcal{B}^2(\pi^0\to\gamma\gamma) \epsilon} \nonumber\\ 
     &= (4.90\pm 0.17_\textnormal{stat.}\pm 0.77_\textnormal{syst.})\times10^{-5},
  \end{align}
  \noindent where the BFs $\mathcal{B}(\omega\to\pi^+\pi^-\pi^0)$ and $\mathcal{B}(\pi^0\to\gamma\gamma)$ are taken from Ref.\,\cite{pdg2016}, $N_{\eta_c}$ is the $\eta_c$ signal yield determined from the best PWA fit, $\epsilon=3.42\%$ is the detection and reconstruction efficiency and $N_{J/\psi}=(1310.6\pm7.0)\times10^6$\,\cite{Ablikim:2016fal} is the number of $J/\psi$ events. Taking into account the measured BF for the $J/\psi\to\gamma\eta_c$ decay, which has large uncertainties, the BF of the $\eta_c$ decay is given by 
  \begin{align}
     \mathcal{B}(\eta_c&\to\omega\omega) \nonumber \\ 
     &= (2.88\pm 0.10_\textnormal{stat.}\pm 0.46_\textnormal{syst.} \pm 0.68_\textnormal{ext.})\times10^{-3}.
  \end{align}
  The last quoted uncertainty corresponds to the error of the $J/\psi\to\gamma\eta_c$ BF and is the dominant uncertainty of this measurement. 
  \section{Mass and Width of the \boldmath{$\eta_c$}}
    The mass and width of the $\eta_c$ are left as free parameters in the PWA fits. The systematic uncertainty of the extracted values is estimated from alternative fits with different fit ranges, different fit hypothesis and the usage of the alternative damping factor. All sources of systematic uncertainties are assumed to be independent and thus their deviations from the nominal result are added in quadrature. The values are found to be 
    \begin{align}
       M(\eta_c) &= (2985.9\pm0.7_\text{stat.}\pm2.1_\text{syst})\,\text{MeV}/c^2 \text{ and } \\
       \Gamma(\eta_c) &= (33.8\pm1.6_\text{stat.}\pm4.1_\text{syst.})\,\text{MeV},
    \end{align}
    where the first uncertainties are statistical and the second systematic. 
    The mass and width are consistent with the world average values.
    
\section{SUMMARY AND DISCUSSION}
Using a sample of $(1310.6\pm7.0)\times10^6$ $J/\psi$ events accumulated with the BESIII detector, we report the first observation of the decay $\eta_c\to\omega\omega$ in the process $J/\psi\to\gamma\omega\omega$. By means of a PWA the branching fraction of $\eta_c\to\omega\omega$ is measured to be $\mathcal{B}(\eta_c\to\omega\omega) = (2.88\pm 0.10_\textnormal{stat.}\pm 0.46_\textnormal{syst.} \pm 0.68_\textnormal{ext.})\times10^{-3}$, where the external uncertainty refers to that arising from the branching fraction of the decay $J/\psi\to\gamma\eta_c$. The obtained value is about one order of magnitude larger than what is expected from NLO perturbative QCD calculations including higher twist contributions. The mass and width of the $\eta_c$ are determined to be $M = (2985.9\pm0.7_\text{stat.}\pm2.1_\text{syst.})\,\text{MeV}/c^2$ and $\Gamma = (33.8\pm1.6_\text{stat.}\pm4.1_\text{syst.})\,\text{MeV}$. The extracted values for the mass and width of the $\eta_c$ are in good agreement with the world average values. This measurement provides new insights into the decay characteristics of charmonium resonances.
\begin{acknowledgments}The BESIII collaboration thanks the staff of BEPCII and the IHEP computing center for their strong support. This work is supported in part by National Key Basic Research Program of China under Contract No. 2015CB856700; National Natural Science Foundation of China (NSFC) under Contracts Nos. 11335008, 11425524, 11625523, 11635010, 11735014; the Chinese Academy of Sciences (CAS) Large-Scale Scientific Facility Program; the CAS Center for Excellence in Particle Physics (CCEPP); Joint Large-Scale Scientific Facility Funds of the NSFC and CAS under Contracts Nos. U1532257, U1532258, U1732263; CAS Key Research Program of Frontier Sciences under Contracts Nos. QYZDJ-SSW-SLH003, QYZDJ-SSW-SLH040; 100 Talents Program of CAS; INPAC and Shanghai Key Laboratory for Particle Physics and Cosmology; German Research Foundation DFG under Contracts Nos. Collaborative Research Center CRC 1044, FOR 2359; Istituto Nazionale di Fisica Nucleare, Italy; Koninklijke Nederlandse Akademie van Wetenschappen (KNAW) under Contract No. 530-4CDP03; Ministry of Development of Turkey under Contract No. DPT2006K-120470; National Science and Technology fund; The Swedish Research Council; U. S. Department of Energy under Contracts Nos. DE-FG02-05ER41374, DE-SC-0010118, DE-SC-0010504, DE-SC-0012069; University of Groningen (RuG) and the Helmholtzzentrum fuer Schwerionenforschung GmbH (GSI), Darmstadt.
\end{acknowledgments}


\bibliography{references}

\end{document}